\documentclass[aps,prb,twocolumn,showpacs,superscriptaddress,floatfix]{revtex4}
\usepackage{graphicx}
\usepackage{bm}

\newcommand{\beq}{\begin{equation}}
\newcommand{\eeq}{\end{equation}}

\newcommand{\av}{\mbox{\scriptsize Av}}
\newcommand{\qcp}{\mbox{\scriptsize QCP}}
\newcommand{\fc}{\mbox{\scriptsize FC}}

\begin{document}
\title{
Flattening of Single-Particle Spectra in Strongly Correlated
Electron Systems \\ and the Violation of the Wiedemann-Franz Law}
\author{V.~A.~Khodel}
\affiliation{Russian Research Centre Kurchatov
Institute, Moscow, 123182, Russia}
\author{V.~M.~Yakovenko}
\affiliation{Condensed Matter Theory Center and Center for
Superconductivity Research, Department of Physics,
University of Maryland, College Park, Maryland 20742-4111, USA}
\author{M.~V.~Zverev}
\affiliation{Russian Research Centre Kurchatov
Institute, Moscow, 123182, Russia}

\date{\today}
\begin{abstract}
The renormalization of the Wiedemann-Franz (WF) ratio
in strongly correlated electron systems is analyzed within the
Landau quasiparticle picture. We demonstrate that the WF law is
violated: (i) at the quantum critical point, where the effective
mass diverges, and (ii) beyond a point of fermion  condensation,
where the single-particle spectrum $\epsilon(p)$ becomes flat.
Results of the analysis are compared with available experimental
data.
\end{abstract}


\maketitle

\paragraph{Introduction.}
In Landau theory,\cite{lan,lanl} a foundation of understanding of
phenomena in condensed matter,  Fermi liquid (FL) is treated
as a system of interacting quasiparticles, having an ideal-Fermi-gas-like
momentum distribution, (hereafter we set the Boltzmann's constant
$k_B=1$),
\beq
n(p)=\left[1+e^{\epsilon(p)/T}\right]^{-1} \ .
\label{dist}
\eeq
In the vicinity of the Fermi surface, the FL single-particle spectrum
$\epsilon(p)$, measured from the chemical potential $\mu$, also has
the ideal-Fermi-gas form
\beq
\epsilon(p)=v_F(p-p_F)\equiv p_F(p-p_F)/M^* \ ,
\label{meff}
\eeq
where the bare mass $M$
is replaced by a $T$-independent effective mass $M^*$.
Therefore FL thermodynamic
and transport properties differ from those of ideal Fermi gas merely
by a numerical factor --- a feature,
inherent in atomic nuclei, liquid $^3$He and conventional metals.

One of the most prominent justifications of the applicability of
the Landau-Migdal quasiparticle picture to metals is associated
with the low-temperature behavior of the Lorenz number $L(T)$, the
ratio of the thermal conductivity $\kappa(T)$ to the product of
temperature $T$ and conductivity $\sigma(T)$. The $T\to 0$ limit
$L_0$, derived by Sommerfeld well before the creation  of FL
theory, equals \cite{kin,ashkroft}
\beq
L_0= \lim_{T\to 0}\ {\kappa(T)\over \sigma(T) T}={\pi^2\over 3e^2} \ .
\label{wfr}
\eeq
Eq.~(\ref{wfr}), known as the Wiedemann-Franz (WF)
law, holds in normal states of  electron
systems of metals,\cite{ott,amato,kambe,syme,ronning} except for
i)  heavy-fermion metals \cite{paschen,paglione,science} CeNiSn
 and CeCoIn$_5$,
ii) an electron-doped material \cite{greene}
Pr$_{2-x}$Ce$_x$CuO$_{4-y}$, and iii) an underdoped compound
\cite{taillefer} YbBa$_2$Cu$_3$O$_y$. In CeNiSn, the experimental
value of the reduced Lorenz number $L(T)/L_0\simeq 1.5$ changes
little at $T<1$\,K that rules out the relevance of phonons to the
violation of the WF law. In the electron-doped compound
Pr$_{2-x}$Ce$_x$CuO$_{4-y}$, the departure of the ratio $L(T)/L_0$
from 1 at $T>0.3$\,K is positive as well and even
larger,\cite{greene} than in CeNiSn.

Predictions of FL theory, including Eq.~(\ref{wfr}), fail in the
vicinity of the so-called quantum critical point (QCP)  where the
effective mass $M^*$ diverges, since at the QCP, the FL spectrum
(\ref{meff}) becomes meaningless. In a standard scenario of the
QCP,\cite{coleman1,coleman2} the divergence of the effective mass
is attributed to vanishing of the quasiparticle weight $z$ in the
single-particle state close to points of second-order phase
transitions, implying that the FL quasiparticle picture of
phenomena breaks down. In dealing with the WF law, a scenario of
its violation, associated with critical fluctuations, is recently
advanced in Ref.~\onlinecite{sachdev}. However, the standard  scenario
of the QCP is flawed,\cite{cond} and therefore in this article,
we employ a different {\it topological} scenario of the QCP, where
the departure of the Lorenz number $L(0)$ from the WF value is
associated with a rearrangement of single-particle degrees of
freedom, a phenomenon described within the Landau quasiparticle
picture.

\paragraph{FL formulas for transport coefficients.}
Within Landau theory, relation between conductivities $\sigma$ and
$\kappa$, and the Seebeck thermoelectric  coefficient $S$, has the
form \cite{kin}
\beq
{\kappa(T)\over \sigma(T) T}+S^2(T)= {1\over e^2}{I_2(T)\over I_0(T)} \ .
\label{wfL}
\eeq
Here
\beq
S(T)={1\over e}{I_1(T)\over I_0(T)}  \  ,
\label{see}
\eeq
and
\beq
I_k(T)= -\int\!\! \left({\epsilon(p)\over T}\right)^k \!\!
   \left({d\epsilon(p)\over dp}\right)^2 \!\!  \tau(\epsilon,T)
   {\partial n(p)\over \partial\epsilon(p)} d\upsilon ,
\label{ik}
\eeq
where $\tau$ is the collision time, $d\upsilon$ is the volume element of
momentum space, and $n(p)$ is given by Eq.~(\ref{dist}).

 Overwhelming contributions to the
integrals $I_k$ come from a narrow vicinity $|\epsilon|\sim T$ of
the Fermi surface. In  Fermi liquids, obeying Landau theory, the Seebeck
coefficient $S(T)$ vanishes linearly with $T$
at $T\to 0$, similarly to the specific heat, given by
\beq
C(T)=-\int \epsilon (p){\partial n(p)\over \partial  T}
d\upsilon \  .
\eeq
Furthermore, in an ideal Fermi gas with impurities, the two quantities
are connected with each other by relation \cite{behnia}
\beq
q(T)={eS(T)N_{\av}\over C(T)}=-1  \   ,
\label{qsc}
\eeq
where $N_{\av}$ is the Avogadro number. In interacting Fermi gases,
the single-particle spectrum $\epsilon(p)$ is not a parabolic
function of $p$, and $q(T)\neq -1$, nevertheless, the
proportionality between  $S(T)$ and $C(T)$ holds, and therefore
the contribution of the Seebeck coefficient to Eq.~(\ref{wfL})
 remains minor.

 In conventional Fermi liquids, the group velocity can be factored out
from the integrals (\ref{ik}). The same is valid
for the collision time $\tau$, depending at $T\to 0$ merely on
impurity scattering. This yields \cite{kin,ashkroft} $I_1(T=0)=0$,
and $I_2(T\to 0)/I_0(T\to 0)=\pi^2/3$. Upon inserting these
numbers into Eq.~(\ref{wfL}) we do arrive at Eq.~(\ref{wfr}) that
holds, even if several bands cross the Fermi surface
simultaneously.

It is worth noting that top-quality samples, used in modern
experiments, possess an extremely low residual resistivity
$\rho_0<0.5\mu\Omega$cm, and often at temperatures, at which
measurements are carried out, electron-electron scattering comes
into play, rendering the FL damping rate $\gamma\sim \tau^{-1}$
$\epsilon$-dependent,  \cite{pines}
\beq
\gamma (\epsilon,T) =
    \gamma(0,0)+\gamma_{ee}T^2(1+\left(\epsilon/2\pi T \right)^2)
    \ ,
\label{tee} \eeq that alters the Lorenz number $L(T)$. To
illustrate the change of its value we neglect the impurity
scattering and set $\gamma(0,0)=0$. In this case, the integrals
$I_k$ \beq I_k= -{1\over \gamma_{ee}T^2}\int\!\!
\left({\epsilon\over T}\right)^k \!\! \left({d\epsilon\over
dp}\right)^2 \! {\partial n(\epsilon(p))\over \partial\epsilon}
{d\upsilon \over 1+\left(\epsilon/2\pi T \right)^2} \label{ikr}
\eeq are calculated numerically, the results are inserted into
Eq.~(\ref{wfL}) to yield \beq L(T)/L_0=0.84\, \ \eeq that agrees
with results obtained in Refs.~\onlinecite{Cat_Al,Catelani} by a
different method, giving rise to a slight decrease of $L(T)$ at
temperatures where the resistivity $\rho(T)$ ceases to be
constant. This suppression explains the departure of the
experimental WF ratio from the FL value (\ref{wfr}), found in the
heavy-fermion metal YbBa$_2$Cu$_3$O$_y$.

\begin{figure}[t]
\includegraphics[width=0.7\linewidth,height=0.54\linewidth]{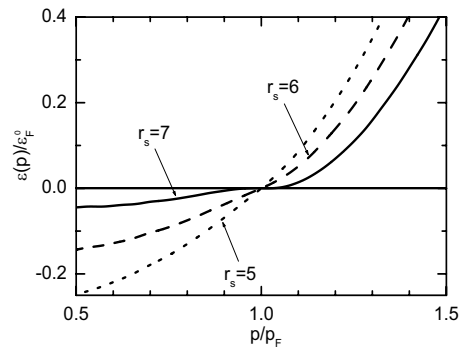}
\caption{ Single-particle spectrum $\epsilon(p)$ of a homogeneous
2D
 electron gas in units of $\epsilon_F^0=p^2_F/2M$,
evaluated \cite{bz} at $T=0$ for different values of the dimensionless
 parameter $r_s=\sqrt{2}Me^2/p_F$.}
\label{fig:2deg}
\end{figure}

\paragraph{Violation of the WF law at the topological quantum critical
point.} In  homogeneous Fermi liquids, the QCP was uncovered first
in experimental studies of a two-dimensional (2D) electron
gas,\cite{shashkin1,pudalov,shashkin2,shashkincond} where the
electron effective mass $M^*(\rho)$ diverges at a critical value
$r_c\simeq 7.0$ of a dimensionless parameter
$r_s=\sqrt{2}Me^2/p_F$. Results \cite{bz} of microscopic
calculations of 2D electron spectra $\epsilon(p,T=0,r_s)$ at
$r_s=5$, 6 and 7 are shown in Fig.~\ref{fig:2deg}. We see that
$\epsilon(p,T=0,r_s)$ becomes flatter and flatter, as  the
critical value $r_s=7.0$  approaches, at which for the first time
the group velocity $v=d\epsilon(p)/dp$ vanishes at the Fermi
surface. As seen from Fig.~\ref{fig:2deg}, the QCP electron
spectrum, denoted further $\epsilon_c(p,T=0)$, has an inflection
point $\epsilon_c(p,T=0)\propto {(p-p_F)^3}$.

%%%%%%%%%%%%%%%%%%%%%%%%%%%%%%%%%%%%%%%%%%%%%%%%%%%%%%%%%%%%%%%%%
%%%%%                Figure: 2deg.eps
%%%%%%%%%%%%%%%%%%%%%%%%%%%%%%%%%%%%%%%%%%%%%%%%%%%%%%%%%%%%%%%%%
%%%%%%%%%%%%%%%%%%%%%%%%%%%%%%%%%%%%%%%%%%%%%%%%%%%%%%%%%%%%%%%%

At finite $T$, the QCP group velocity $d\epsilon_c(p,T)/dp$
acquires a  finite value,\cite{shag} and the QCP single-particle
spectrum becomes \cite{ckz}
\beq
\epsilon_c(p,T)=p_F{p-p_F\over M^*(T)}+{v_2\over 3}(p-p_F)^3 \  .
\label{eqcp}
\eeq
The $T$-dependence of the effective mass $M^*(T)$ is evaluated
on the base of the Landau equation \cite{lan,lanl}
\beq
{\partial \epsilon( p)\over \partial  p}={p\over M}
+{1\over 3}\int f_1( p, p_1)
{\partial n( p_1)\over \partial p_1}{p^2_1dp_1\over \pi^2} \ ,
\label{lansp}
\eeq
connecting  $\epsilon(p)$ with the quasiparticle momentum
distribution $n(p)$, given by Eq.~(\ref{dist}), in terms of the
first harmonic $f_1$ of the interaction function $f$. If  $f(q)$
is an analytical function of $q$, leading $T$-dependent
contributions to $\epsilon(p,T)$ come from first terms of Taylor
expansion of $f(p,p_1)$ vs $(p-p_F)$ and $(p_1-p_F)$. As a result,
one obtains \cite{ckz}
\beq
{M\over M^*(T)}=\left({9v_2\over 8}\right)^{1/3}
\left({TM\over p^2_F}\right)^{2/3} \ .
\eeq
To evaluate the integrals $I_k$, we introduce a new integration
variable $\epsilon(p)$ instead of $p$, then express $p-p_F$ in
terms of energy $\epsilon$ from Eq.~(\ref{eqcp}) to find
\beq
y/y_0=\left({\epsilon\over 2}+\sqrt{{\epsilon^2\over 4}
  +{T^2\over 8}}\right)^{\!\!\!1/3}\!\!\!-
   \left(\sqrt{{\epsilon^2\over 4}+{T^2\over 8}}-
{\epsilon\over 2}\right)^{\!\!\!1/3}\!\!\!\!,
\eeq
where $y=p-p_F$ and $y_0=(3Mp_F/v_2)^{1/3}$. With these results,
the integrals $I_k$ are calculated numerically that yields
\beq
L_{\qcp}(0)/L_0=1.81  \  .
\label{wfqcp}
\eeq
Thus close to the QCP, the ratio $L_{\qcp}(0)/L_0$ is enhanced as
compared with the FL value (\ref{wfr}). This result is in
agreement with the experimental value of the violation of the WF
law, observed in Refs.~\onlinecite{paschen,greene}.

\begin{figure}[t]
\includegraphics[width=1.\linewidth,height=1.3\linewidth]{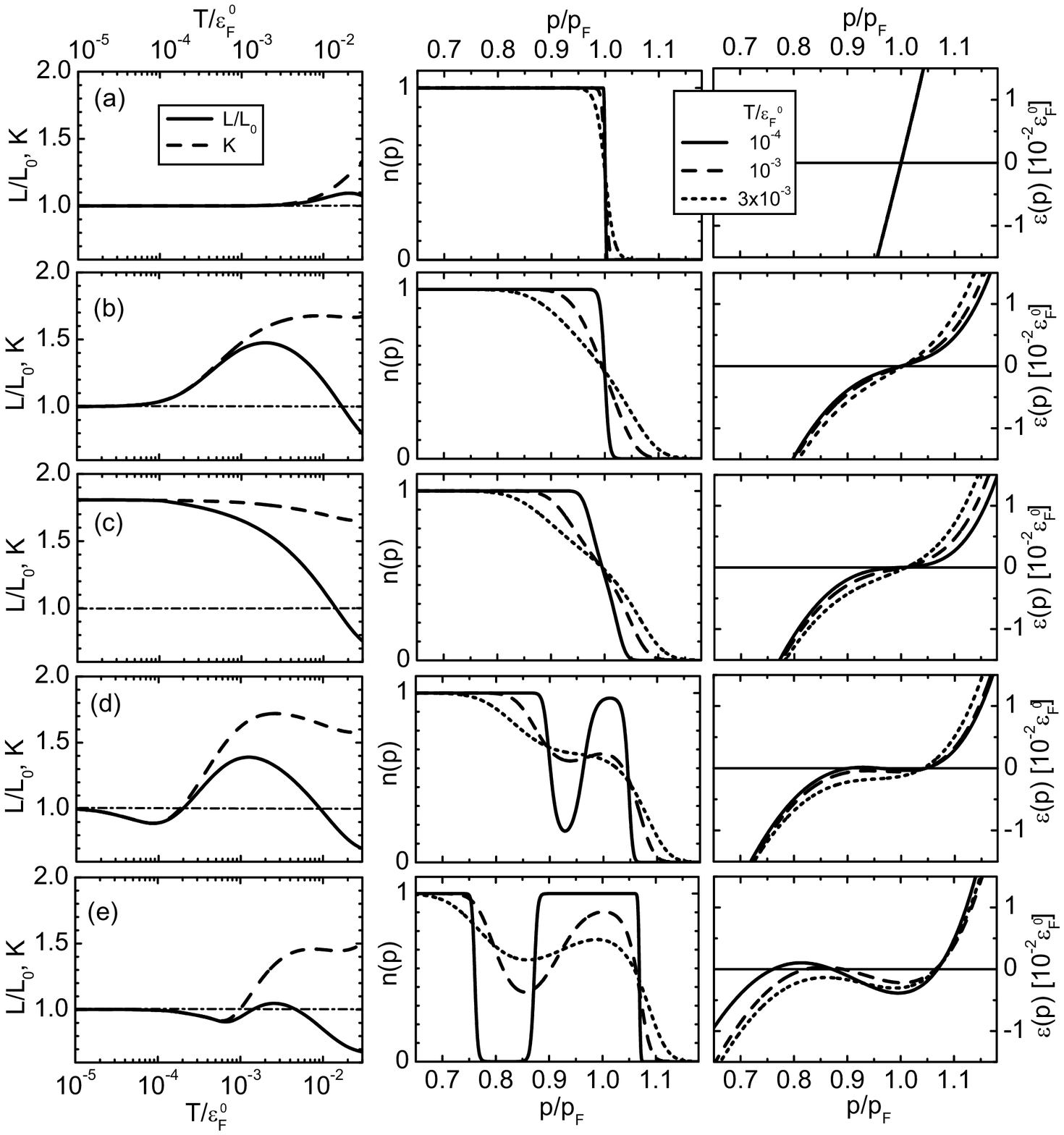}
\caption{ The reduced Lorenz number $L(T)/L_0$ together with the
ratio $K(T)={3I_2(T)/(\pi^2 I_0(T))}$ (left panels) as functions
of the reduced temperature $T/\epsilon_F^0$, the occupation
numbers $n(p)$ (middle panels) and the single-particle spectrum
(right panels) calculated at different line-type-coded
temperatures with the interaction function (\ref{model}) for five
different sets of input parameters, which correspond to five
different cases: Fermi liquid with the effective mass $M^*/M\simeq
6$, i.e.\ quite far from the QCP (panels a); Fermi liquid with
$M^*/M\simeq 80$, which is close to the QCP (panels b); system at
the QCP (panels c), the state just beyond the QCP, with
$p_2-p_1\simeq 0.14\,p_F$ (panels d) and the state well beyond the
QCP, with $p_2-p_1\simeq 0.31\,p_F$ (panels e).}
\label{fig:wf_law}
\end{figure}

\paragraph{Numerical calculations of the Lorenz number $L(T)$ through
the topological QCP.}
In this paragraph we show results of numerical calculations of
Eq.~(\ref{lansp}) with an interaction function $f$, having the analytical
form
\beq
f(q)={\lambda \over (q^2/4p_F^2-1)^2+\alpha^2}\  ,
\label{model}
\eeq
where the parameter $\lambda$ is fixed, while the parameter
$\alpha$, depending on $r_s$, is chosen to provide the best
agreement with the microscopic spectrum, drawn in
Fig.~\ref{fig:2deg}. Results of  calculations of the momentum
distribution $n(p,T)$ and spectrum $\epsilon(p,T)$ at different
values of $\alpha$ are given in the second and third columns of
Fig.~\ref{fig:wf_law} respectively. On the FL side, (the upper
panels (a) and (b)), of the QCP, located at $\alpha=0.479$, the
spectrum $\epsilon(p,T)$ has the standard FL form (\ref{meff})
that holds until the effective mass $M^*$ attains values $\simeq
10^2M$.
 The QCP
spectrum $\epsilon_c(p,T)$, with the group velocity $v_F(T)$
vanishing at $T=0$ is shown in the third panel (c). It differs
from the FL one in two aspects. First, at $T\to 0$ it has the
inflection point. Second, its form drastically changes with $T$
elevation.

%%%%%%%%%%%%%%%%%%%%%%%%%%%%%%%%%%%%%%%%%%%%%%%%%%%%%%%%%%%%%%%%%
%%%%%                Figure: wf_law.eps
%%%%%%%%%%%%%%%%%%%%%%%%%%%%%%%%%%%%%%%%%%%%%%%%%%%%%%%%%%%%%%%%%
%%%%%%%%%%%%%%%%%%%%%%%%%%%%%%%%%%%%%%%%%%%%%%%%%%%%%%%%%%%%%%%%%

Having at hand the spectrum $\epsilon(p)$, the transport integrals
(\ref{ik}) are straightforwardly evaluated. Results of
calculations of the functions $L(T)$ and
$K(T)=3I_2(T)/(\pi^2I_0(T))$ are shown in the first column of
Fig.~\ref{fig:wf_law}. The upper panel (a) of
Fig.~\ref{fig:wf_law} illustrates the situation on the FL side of
the QCP ($\alpha=0.520$, $M^*/M=6$). We observe  no deviations from
the FL predictions. As seen from the panel (b), the WF law also
holds in the immediate vicinity of the QCP ($\alpha=0.482,
M^*/M=80$). However, the departure from FL theory, associated with
the temperature dependence of the spectrum  $\epsilon(p,T)$,
becomes well pronounced in both the ratios $K=3I_2/(\pi^2I_0)$ and
$S=I_1/I_0$ already at extremely low $T\simeq
10^{-3}\epsilon^0_F$. The QCP results are shown in the middle
panel (c). We see that the WF law is, indeed, violated: the ratio
$L(0)/L_0$ turns out to be in excess of 1, in agreement with the
above result (\ref{wfqcp}).

Results, shown in two lower panels (d) and (e), where
$\alpha=0.475$ and $\alpha=0.472$ correspondingly, demonstrate that
as the system goes away from the  QCP, the value of the  group velocity
$v_F(T=0)$ turns out to be finite again, and the WF law (\ref{wfr})
is recovered. However, already at  $T\simeq 10^{-4}\epsilon^0_F$,
even lower, than   on the FL side, the
Lorenz number $L(T)$ becomes $T-$dependent.

\paragraph{Topological phase transitions in strongly correlated
Fermi systems.} As seen from the panels (d) and (e), beyond the
QCP, equation
 \beq
 \epsilon(p)=\mu
 \label{bif}
 \eeq
has three roots $p_1<p_2<p_3$, the curve $\epsilon(p,T=0)$ crosses
the Fermi level three times, and occupation numbers $n(p,T=0)$
become: $n(p)=1$ at $p<p_1$, while at $p_1<p<p_2$, $n(p)=0$; at
$p_2<p<p_3$, once again $n(p)=1$, and at $p>p_3$, $n(p)=0$. Thus
at the coupling constant $g>g_{\qcp}$, the Fermi surface becomes
multi-connected. This is a typical topological phase transition,
at which no one symmetry, inherent in the ground state, is
violated.\cite{zb,arshag,vosk,schofield}  As the interaction
strength  increases, the number of the points, where the spectrum
$\epsilon(p,T=0)$ crosses the Fermi level, rapidly grows, however,
the number of roots of Eq.~(\ref{bif}) remains countable.

In another type of the topological transitions, the so-called fermion
condensation,\cite{ks,vol,noz,physrep,yak,volrev} the roots of
Eq.~(\ref{bif}) form an uncountable set. Indeed, the ground state
energy $E$ is a functional \cite{lan,lanl} of the quasiparticle momentum
distribution $n(p)$, confined within the interval $0<n<1$.  In the strong
coupling limit, the minimum of this functional is found from the
variational condition \cite{ks}
\beq
{\delta E\over \delta n(p)}=\mu    , \quad p \in {\cal C}  \  ,
\label{var}
\eeq
the chemical potential $\mu$, being determined by the requirement
$\sum n(p)=\rho$. Since the l.h.s.\ of Eq.~(\ref{var}) is nothing
but the quasiparticle energy $\epsilon(p)$, we see that in the
case at issue, solutions of Eq.~(\ref{bif}), called the fermion
condensate (FC), exist in a whole domain ${\cal C}$, giving rise
to swelling of the Fermi surface. Remarkably, the presence of the
FC results in breaking of the {\it particle-hole symmetry},
another salient feature of the phenomenon of fermion condensation,
that exhibits itself in a marked violation of the WF law.

True, if the interaction strength is small, the solutions of
Eq.~(\ref{var})  do not meet the restriction $n(p)\leq 1$.
However, beginning with a critical constant, i.e.\ at
$g>g_{\fc}$, these solutions, {\it smooth} functions $n_*(p)$,
meet this restriction wherever. Thus on the Lifshitz phase
diagram, the FL phase occupies a region $g<g_{\qcp}$, a domain
$g_{\qcp}<g<g_{\fc}$ is occupied by the phase, having the
multi-connected Fermi surface, while a domain $g>g_{\fc}$, by the
FC phase.

\paragraph{Breaking of the particle-hole symmetry and the WF law in
systems with a fermion condensate.}

%%%%%%%%%%%%%%%%%%%%%%%%%%%%%%%%%%%%%%%%%%%%%%%%%%%%%%%%%%%%%%%%%
%%%%%                Figure: wf_yu_03.eps
%%%%%%%%%%%%%%%%%%%%%%%%%%%%%%%%%%%%%%%%%%%%%%%%%%%%%%%%%%%%%%%%%
\begin{figure}[ht]
\includegraphics[width=1.\linewidth,height=1.\linewidth]{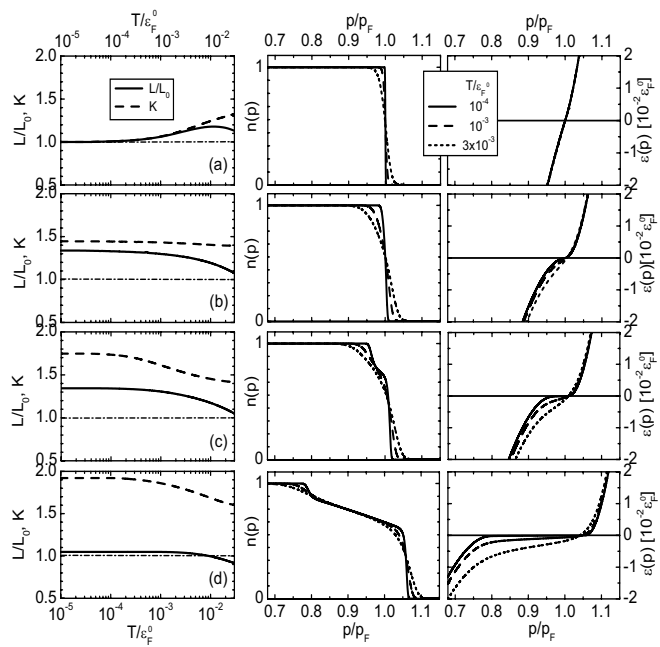}
\caption{ Same as in Fig.~\ref{fig:wf_law} calculated with the
interaction function (\ref{yuk}) for $\beta p_F=3$ and four values
of the parameter $g$ in units of $M^{-1}$: 3.0 (Fermi liquid,
panels a), 3.55 (QCP, panels b), 3.9 (the state with the FC
fraction $\rho_{\fc}/\rho=0.1$, and 4.8 (the state with
$\rho_{\fc}/\rho=0.5$, panels d).} \label{fig:wf_yu_03}
\end{figure}
%%%%%%%%%%%%%%%%%%%%%%%%%%%%%%%%%%%%%%%%%%%%%%%%%%%%%%%%%%%%%%%%%

\begin{figure}[ht]
\includegraphics[width=1.\linewidth,height=1.\linewidth]{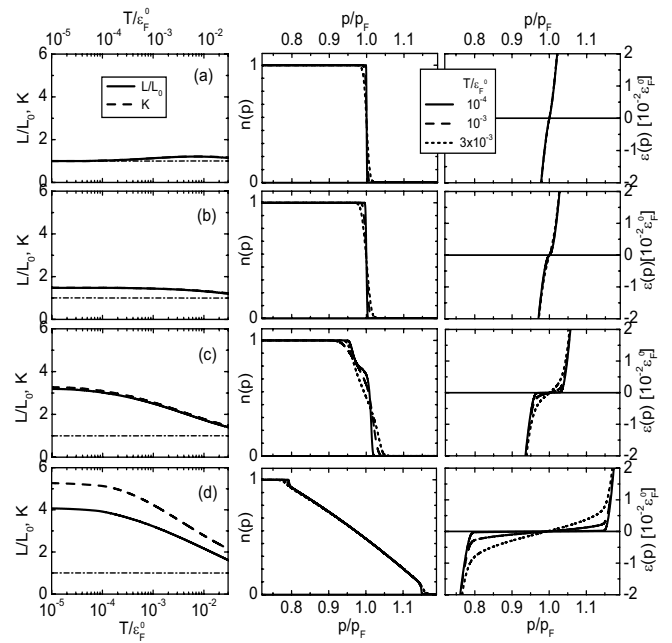}
\caption{ Same as in Fig.~\ref{fig:wf_yu_03} calculated for $\beta
p_F=30$ and four values of the parameter $g$ in units of $M^{-1}$:
25.0 (Fermi liquid, panels a), 33.4 (QCP, panels b), 69.0 (the
state with $\rho_{\fc}/\rho=0.1$, panels c), and 224.0 (the state
with $\rho_{\fc}/\rho=0.5$, panels d).} \label{fig:wf_yu_30}
\end{figure}

\begin{figure}[t]
\includegraphics[width=0.75\linewidth,height=0.8\linewidth]{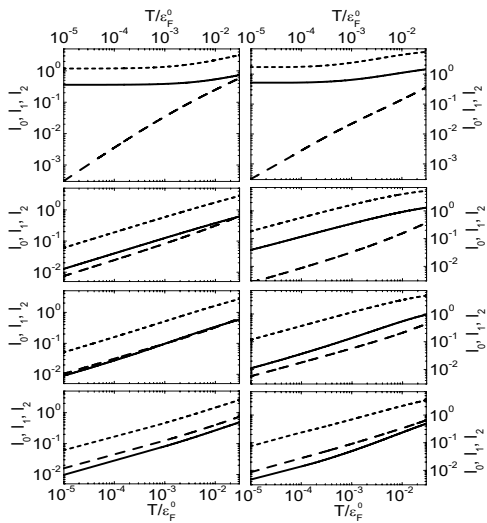}
\caption{ The transport integrals $I_0$ (solid lines), $I_1$
(dashed lines) and $I_2$ (short-dashed lines) in log-log scale as
functions of reduced temperature $T/\epsilon_F^0$, calculated with
the interaction function (\ref{yuk}) for $\beta p_F=3$ (left
column) and $\beta p_F=30$ (right column) and four values of the
parameter $g$ corresponding to FL (upper panels), the QCP (second
line of panels), and to the states with 10\% (third line) and 50\%
(lower panels) of quasiparticles in the FC.} \label{fig:ik_yuu}
\end{figure}

It is worth noting that if the interaction function $f(q)$ possesses
a singularity at $q=0$, then the Lifshitz phase diagram, constructed
above, alters, since the phase with the multi-connected Fermi surface
disappears, and at $g>g_{\qcp}$ the system contains the normal quasiparticles
and a FC fraction that grows linearly with the difference $g-g_{\qcp}$.
This situation is convenient for the demonstration of the impact of
breaking of the particle-hole symmetry in systems with a FC on
the violation of the WF law. In what follows we address
a model with the interaction function
\beq
f(q)=g \exp(-\beta q)/q  \  .
\label{yuk}
\eeq
 The structure of the FC, emerging at
$g>g_{\qcp}$, is  found with the help of formulas, given
in Ref.~\onlinecite{physrep}. It is shown in Figs.~\ref{fig:wf_yu_03} and
\ref{fig:wf_yu_30}. As seen from
Figs.~\ref{fig:wf_yu_03} and \ref{fig:wf_yu_30}, the model
derivative $dn(p)/dp$ has peaks at both the boundary
points of the FC region. Notably vicinities of these points contribute
overwhelmingly to the transport integrals $I_k$, since the
 single-particle spectrum $\epsilon(p,T=0)$ identically
vanishes inside the FC domain.   Outside this domain,\cite{physrep}
\beq
{d\epsilon(p,T=0)\over dp}\propto \sqrt{|\epsilon(p)|} \ .
\label{gryuk}
\eeq
This power behavior results in a marked violation of the WF law.
Indeed, upon inserting Eq.~(\ref{gryuk}) into Eq.~(\ref{ik}) one
finds that major contributions to every of the integrals $I_k$,
proportional to $T^{1/2}$, come from the exterior of the FC
domain, while its interior ensures minor ones, proportional to
$T$.
 This conclusion is confirmed by results of  numerical calculations, shown in
 Fig.~\ref{fig:ik_yuu}

%%%%%%%%%%%%%%%%%%%%%%%%%%%%%%%%%%%%%%%%%%%%%%%%%%%%%%%%%%%%%%%%%
%%%%%                Figure: wf_yu_30.eps
%%%%%%%%%%%%%%%%%%%%%%%%%%%%%%%%%%%%%%%%%%%%%%%%%%%%%%%%%%%%%%%%%
%%%%%%%%%%%%%%%%%%%%%%%%%%%%%%%%%%%%%%%%%%%%%%%%%%%%%%%%%%%%%%%%%

%%%%%%%%%%%%%%%%%%%%%%%%%%%%%%%%%%%%%%%%%%%%%%%%%%%%%%%%%%%%%%%%%
%%%%%                Figure: ik_yuu.eps
%%%%%%%%%%%%%%%%%%%%%%%%%%%%%%%%%%%%%%%%%%%%%%%%%%%%%%%%%%%%%%%%%
%%%%%%%%%%%%%%%%%%%%%%%%%%%%%%%%%%%%%%%%%%%%%%%%%%%%%%%%%%%%%%%%%

The transport integrals $I_k$ are calculated numerically. In
Fig.~\ref{fig:ik_yuu}, we show results of these calculations,
performed with the value of the dimensionless parameters $\beta
p_F=3$ and $\beta p_F=30$. As seen from this figure, at the QCP
and beyond it the integrals $I_0(T=0)$ and $I_1(T=0)$ has the same
order, implying that in contrast to FL theory, the $T=0$ value of
the Seebeck coefficient $S(0)=I_1(0)/eI_0(0)$ differs from 0, an
exhibition of breaking of the particle-hole symmetry, occurring in
any system with a FC.  On the other hand, the ratio $L(0)/L_0$
turns out to be even larger than at the QCP. Furthermore,
calculations demonstrate that  with increasing $\beta$, the value
of this ratio increases as well, i.e. the longer the radius of the
interaction function (\ref{yuk}) in the coordinate space, the
larger is the departure from the WF law.

\paragraph{Anisotropy of the violation of the WF law close to the QCP in
heavy-fermion metals.} The anisotropic Fermi surface of the
majority of heavy-fermion metals has a sector, where at $T\to 0$,
the quasiparticle group velocity $v_F$ keeps a finite value,
implying that the WF law holds. However, recently in experimental
studies of CeCoIn$_5$ in external magnetic fields, close to the
critical value $H_c$, suppressing superconductivity of this metal,
the WF law was found to be violated.\cite{science} The violation
is anisotropic that cannot be attributed to the collapse of
collective degrees of freedom. On the other hand, close to the
topological QCP, the conductivity tensors $\sigma_{ik}\propto
{<}v_iv_k{>}$ and $\kappa_{ik}\propto {<}\epsilon({\bf
p})v_iv_k{>}$ become anisotropic, and this anisotropy is well
pronounced  in sufficiently large magnetic fields. Indeed, the
magnetic field does not affect the $z-$components of the velocity
${\bf v}$, parallel to its direction. As a result, the particular
QCP $T$-dependence of the integrals $I_k$ holds, triggering the
violation of the WF relation
$L_{zz}=\sigma_{zz}/(T\kappa_{zz})=\pi^2/3e^2$. On the other hand,
the electron motion is completely rearranged in the direction,
perpendicular to ${\bf H}$ that leads to a considerable increase
of the respective components of the group velocity and the
suppression of the departure of the respective components of the
ratio ${\bf L}$ from their WF value.

\paragraph{Conclusion.}
In conclusion, we have demonstrated that flattening of
single-particle spectra $\epsilon(p)$ of strongly correlated
electron systems considerably changes their transport properties,
especially beyond the point of fermion condensation due to
breaking of the particle-hole symmetry.
Results of our analysis demonstrate that search for the violation
of the WF law in new materials should be confined to electron
systems without disorder, possessing, nevertheless, a sufficiently
large resistivity, the fact, justifying a minor role of light
carriers. One more distinguished feature of systems, where marked
departures from  the WF law can exist, is related to the
enhancement  of the Seebeck coefficient $S(T\to 0)$ and the ratio
$q(T\to 0)=eS(T\to 0)/C(T\to 0)$. Interestingly, these features
are inherent in the heavy--fermion metals where the  violation of the WF
law was observed. Indeed, in CeNiSn, the $q$ value exceeds
 $10^2$.\cite{behnia} Its value is markedly enhanced in compounds
Pr$_{2-x}$Ce$_x$CuO$_{4-y}$ as well.\cite{li}

We gratefully acknowledge discussions with J.~Paglione,
F.~Steglich, L.~Taillefer and M.~A.~Tanatar. This research was
supported by the McDonnell Center for the Space Sciences, by Grant
No.~NS-8756.2006.2 from the Russian Ministry of Education and
Science, and by Grants Nos.~06-02-17171 and 07-02-00553 from the
Russian Foundation for Basic Research.

\end{document}